\documentclass[aps,prl,twocolumn,superscriptaddress]{revtex4}
\usepackage{mathrsfs}
\usepackage{epsfig}
\usepackage{graphicx}
\usepackage{amsfonts}
\usepackage[figuresright]{rotating}
\usepackage{amssymb}
\usepackage{amsmath}
\usepackage{dcolumn}
\usepackage{bm}
\usepackage{color}

\def\be{\begin{equation}} \def\ee{\end{equation}}
\def\bea{\begin{eqnarray}} \def\eea{\end{eqnarray}}

\newcommand{\WQC} {Wilczek Quantum Center and Key Laboratory of Artificial Structures and Quantum Control, School of Physics and Astronomy, Shanghai Jiao Tong University, Shanghai 200240, China}

\newcommand{\SRCQC}{Shanghai Research Center for Quantum Sciences, Shanghai 201315, China}

\begin{document}
\title{Nonergodic delocalized paramagnetic states in quantum neural networks}

\author{Shuohang Wu}
\affiliation{\WQC}


\author{Zi Cai}
\email{zcai@sjtu.edu.cn}
\affiliation{\WQC}
\affiliation{\SRCQC}

\begin{abstract}  
Typically, it is assumed that a high-energy eigenstate of a generic interacting quantum many-body Hamiltonian is thermal and obeys the eigenstate thermalization hypothesis. In this work, we show that the paramagnetic phase of a quantum Hopfield neural network model is  delocalized but nonergodic. The combination of permutational symmetry and frustration in this model organize its high-energy eigenstates  into clusters, which can each be considered a large quantum spin and has no correlation with others. This model provides another ergodicity-breaking mechanism in quantum many-body systems.

\end{abstract}


\maketitle

{\it Introduction --} For a generic (closed) many-body system, it is postulated that its trajectory  will equiprobably cover the constant energy manifold in phase space on long timescales (ergodicity hypothesis),  which justifies the statistical ensemble description of the macroscopic quantities of equilibrium systems\cite{Rigol2008}.  Two prototypical examples of ergodicity breaking are the many-body localization(MBL)\cite{Basko2006,Oganesyan2007,Znidaric2008,Pal2010} and spin glass(SG)\cite{Mezard1986}. The physical origin of ergodicity breaking in MBL is the quantum-interference effect, whereas that in  SG is due to frustration and randomness. Other examples include integrable systems with extensive conserved quantities\cite{Rigol2007}, and quantum systems with special eigenstates (dubbed ``quantum scarred'' states\cite{Turner2018}). Searching for nonergodic systems and nonthermalized dynamics is not only of fundamental interest in statistical physics, but also of immense potential significance in quantum information processing.

An SG  is a disordered magnetic system whose local magnetic moments are frozen along random orientations at low temperatures\cite{Edwards1975}, leading to a ``magnetic'' phase without conventional long-range order. Although the SG is essentially classical, incorporating the quantum effect gives rise to a plethora of novel phenomena due to the interplay between quantum fluctuations and the frustration-induced rugged energy landscape\cite{Baldwin2017,Mukherjee2018,Rademaker2020,Thomson2020,Winer2022}. For instance, in a quantum p-spin model, it is proposed that the SG phase breaks the ergodicity yet is not MBL\cite{Baldwin2017}, whereas numerical studies of a quantum Sherrington and Kirkpatrick model  seem to suggest a different scenario\cite{Mukherjee2018}. The study of SG is interesting not only in the context of statistical mechanics, but also in the context of memory models. For instance, the Hopfield neural network (HNN) was introduced as a toy model of associative memory\cite{Hopfield1982}, where the memory patterns are to be retrieved via classical annealing\cite{Amit1985a,Amit1985b}. The strong disorder correlation between the bonds in the HNN yields a structure considerably simpler\cite{Hemmen1982} than that in conventional SG models.

In this study, we investigate a quantum generalization of the HNN model, where the intrinsic quantum fluctuations are introduced  by implementing a transverse magnetic field\cite{Rotondo2018,Fiorelli2020}. Unlike standard discussions of quantum SG or HNN models, our focus is not on the low energy states (e.g. the SG or retrieval phases), but is on the high-energy paramagnetic states. It is widely believed that the high-energy eigenstates of a generic many-body Hamiltonian  satisfy the eigenstate thermalization hypothesis\cite{Deutsch1991,Srednicki1994}(ETH), wherein the reduced density matrix of a subsystem in a given eigenstate equals  the microcanonical or canonical  ensemble description set by the energy density of the eigenstate, because the remainder of the system acts as a heat bath thus thermalizing the subsystem of interest. It applies for a large class of quantum many-body systems, including systems with conventional ({\it e.g.}, ferromagnetism\cite{Fratus2015}) or unconventional (SG\cite{Mukherjee2018}) symmetry-breaking phases, as well as MBL systems with many-body mobility edges\cite{Laumann2014}. Despite the significant discrepancy between the low-energy states of these models, their high-energy eigenstates share  common properties and obey the ETH. However, it is shown herein that the paramagnetic eigenstates of the quantum HNN Hamiltonian violates  the ETH, but is not MBL either, which can be understood as a consequence of spontaneous clustering.

 \begin{figure}[htb]
\includegraphics[width=0.99\linewidth]{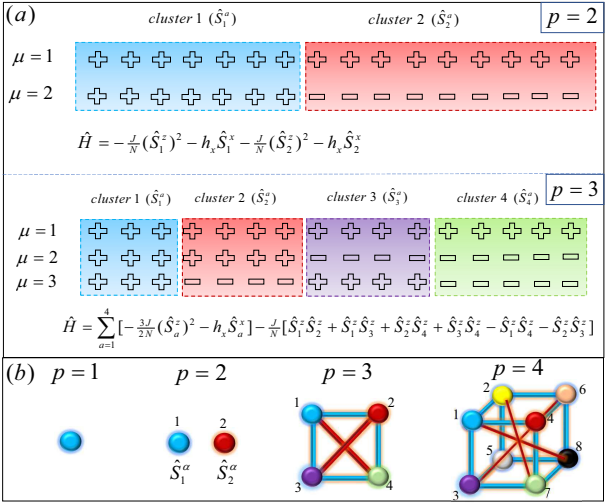}
\caption{(Color online)(a) Classification scheme of lattice sites and effective Hamiltonian for  $p=2$ and $p=3$ (b) Sketches of the effective Hamiltonian for  $p=1\sim 4$, where the large spins are located on the vertex, and the blue(red) bonds represent FM (AFM) couplings between them.} \label{fig:fig1}
\end{figure}

{\it Model and method --} The studied model is a quantum generalization of the HNN model whose Hamiltonian is a transverse Ising model with all-to-all coupling:
\begin{equation}
H= -\frac 12\sum_{i\neq j} J_{ij}\hat{s}_i^z \hat{s}_j^z-h_x \sum_i \hat{s}_i^x \label{eq:Ham1}
\end{equation}
where $\hat{s}_i^{\alpha}=\frac 12 \hat{\sigma}_i^{\alpha}$ with $\alpha=x,y,z$ and $\hat{\sigma}_i^\alpha$ are Pauli matrices on site i, and $h_x$ is the strength of the uniform transverse magnetic field. The interaction strength between sites i and j is defined as:
\begin{equation}
J_{ij}=\frac JN \sum_{\mu=1}^p \xi_i^\mu \xi_j^\mu \label{eq:coupling}
\end{equation}
where $N$ is the number of lattice sites. $p$ is the number of patterns embedded in the system ($\mu$ is the  pattern index), where each pattern can be considered as a N-dimensional vector $\vec{\xi}^\mu=\{ \xi_1^\mu, \xi_2^\mu, \cdots, \xi_N^\mu \}$  with $\xi_i^\mu$ taken to be quenched, independent, random variables ($\xi_i^\mu=\pm 1$ with equal probabilities). Memory patterns $\vec{\xi}^\mu$  are stored in the quenched random couplings via Eq.(\ref{eq:coupling}).

In the absence of the transverse field ($h_x=0$), the Hamiltonian.(\ref{eq:Ham1}) is reduced to a classical HNN model with the capacity of retrieval of information embedded in the memory patterns $\vec{\xi}^\mu$. If one begins from a classical spin configuration similar enough to one of the stored  patterns $\vec{\xi}^\mu$, the HNN system could retrieve the correct pattern via classical annealing. In the thermodynamic limit ($N\rightarrow\infty$), such a retrieval occurs if $p/N$ is less than a critical value\cite{Amit1985a}, whereas for cases with a finite  p, the energy of the classical Hamiltonian is minimized by the 2p spin configurations: $\vec{s}=\pm \frac 12\vec{\xi}^{1\sim p}$  (Mattis states).  The symmetric and asymmetric mixing of these Mattis states as metastable states have also been analyzed\cite{Amit1985b}.

Now, we turn to the quantum Hamiltonian.(\ref{eq:Ham1}) ($h_x>0$), where  we choose the basis as the eigenstates of $\hat{s}^z$: $|\vec{s}\rangle=|s_1^z,\cdots,s_N^z\rangle$. The analysis in this work is restricted to the case of a small finite  p. We first analyze the symmetry of the Eq.(\ref{eq:Ham1}). The simplest case is $p=1$, where the disorder can be gauged away\cite{Mattis1976}, and there is no frustration. By performing a gauge transformation: $\tilde{s}_z^i=\xi_i^1\hat{s}^z_i$, the Hamiltonian.(\ref{eq:Ham1}) become an FM transverse Ising model with uniform all-to-all coupling, where the $\bf{permutational}$ ${\bf symmetry}$ (PS) among different sites allows us to combine all the spins into a large spin with  operators $\hat{S}^\alpha=\sum_i \tilde{s}^\alpha_i$\cite{Carollo2021} and the Eq.(\ref{eq:Ham1}) becomes a Lipkin-Meshkov-Glick(LMG) Hamiltonian\cite{Lipkin1965}:
\begin{equation}
H_1=-\frac J{2N} (\hat{S}^z)^2- h_x \hat{S}^x.
 \end{equation}
Such a gauge transformation applies  not only  for $p=1$, but also for general $p$ cases, where we can always choose one of the patterns ({\it e.g.}pattern 1), and transform it into an FM pattern via the transformation defined above (other patterns are also changed accordingly). Therefore, without losing generality, in the following discussion, we always choose pattern 1 as the FM pattern($\xi_i^1=1$ $\forall$ i).

For $p=2$, the system can be divided into two clusters according to the sign of $\xi_i^1 \xi_i^2$:  the $i$th site satisfying $\xi_i^1 \xi_i^2=1$($-1$) belongs to cluster 1 (2). Similar to $p=1$, the PS  within each cluster enables us to combine the spins within it as: $\hat{S}^\alpha_a=\sum_{i\in a} \hat{s}^\alpha_i$ where $a=1,2$ is the cluster index. It is easy to check there is no coupling between the two clusters according to Eq.(\ref{eq:coupling}), and the Hamiltonian turns to two decoupled LMG models, as shown in Fig.\ref{fig:fig1}.   For the cases with $p>2$,  the lattice sites can be classified into $2^{p-1}$ clusters, each of which is a large spin interacting with others via the ferromagnetic (FM) or antiferromagnetic (AFM) coupling. For example, the classification scheme of lattice sites for $p=3$ is shown in Fig.\ref{fig:fig1}, where each cluster is represented by a large spin located on the vertex of a square. The blue bonds denote  FM interactions, and the red ones represent AFM couplings, which lead to frustration.

 \begin{figure}[htb]
\includegraphics[width=0.9\linewidth,bb=105 53 761 547]{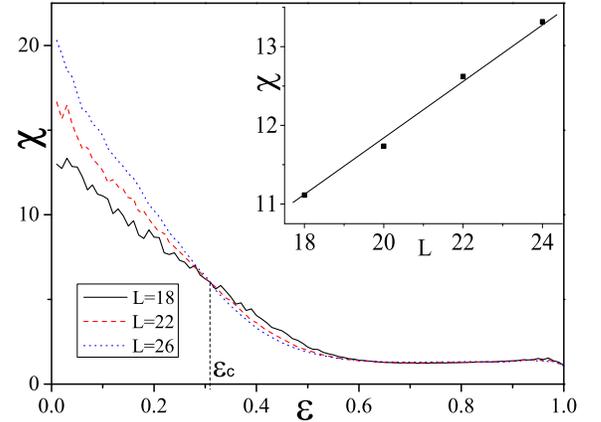}
\caption{(Color online) Spin glass order parameter $\chi$ as a function of normalized energy $\epsilon$ for different system size with $h_x=0.1J$, $\mathcal{N}=1000$.  The inset indicates the system size dependence of $\chi$ with a fixed $\epsilon=0.1$ within the SG phase.} \label{fig:fig2}
\end{figure}

In this work, we study the properties of highly excited eigenstates of the Hamiltonian.(\ref{eq:Ham1}) via the exact diagonalization method. The PS within each cluster allows us to block diagonalize the Hamiltonian. Throughout the paper, we choose the fully symmetric subspace, which corresponds to the Hilbert space with the largest total spin. Accounting for the PS  not only significantly reduces the Hilbert space dimension, but also allows us to resolve the accidental degeneracy between the energy levels in subspaces with different conserved quantities, which is important to analyze the level space statistics. In the following, we focus on the case of $p=4$, which represents a generic situation of the HNN model with finite $p$,   in contrast to the ``special" cases ({\it e.g.},  $p=1,2$). The PS within each of the $2^{4-1}=8$ clusters allows us to derive an effective Hamiltonian represented by a cube with FM and AFM bonds(see the Supplementary material (SM)\cite{Supplementary}). We randomly sample $\mathcal{N}$ sets of independent memory patterns $\{\vec{\xi}^1\cdots\vec{\xi}^p\}$ with $\mathcal{N}=10^3$, and the ensemble average is performed over all $\mathcal{N}$ pattern realizations.

\begin{figure*}[htb]
\includegraphics[width=0.325\linewidth,bb=95 53 760 555]{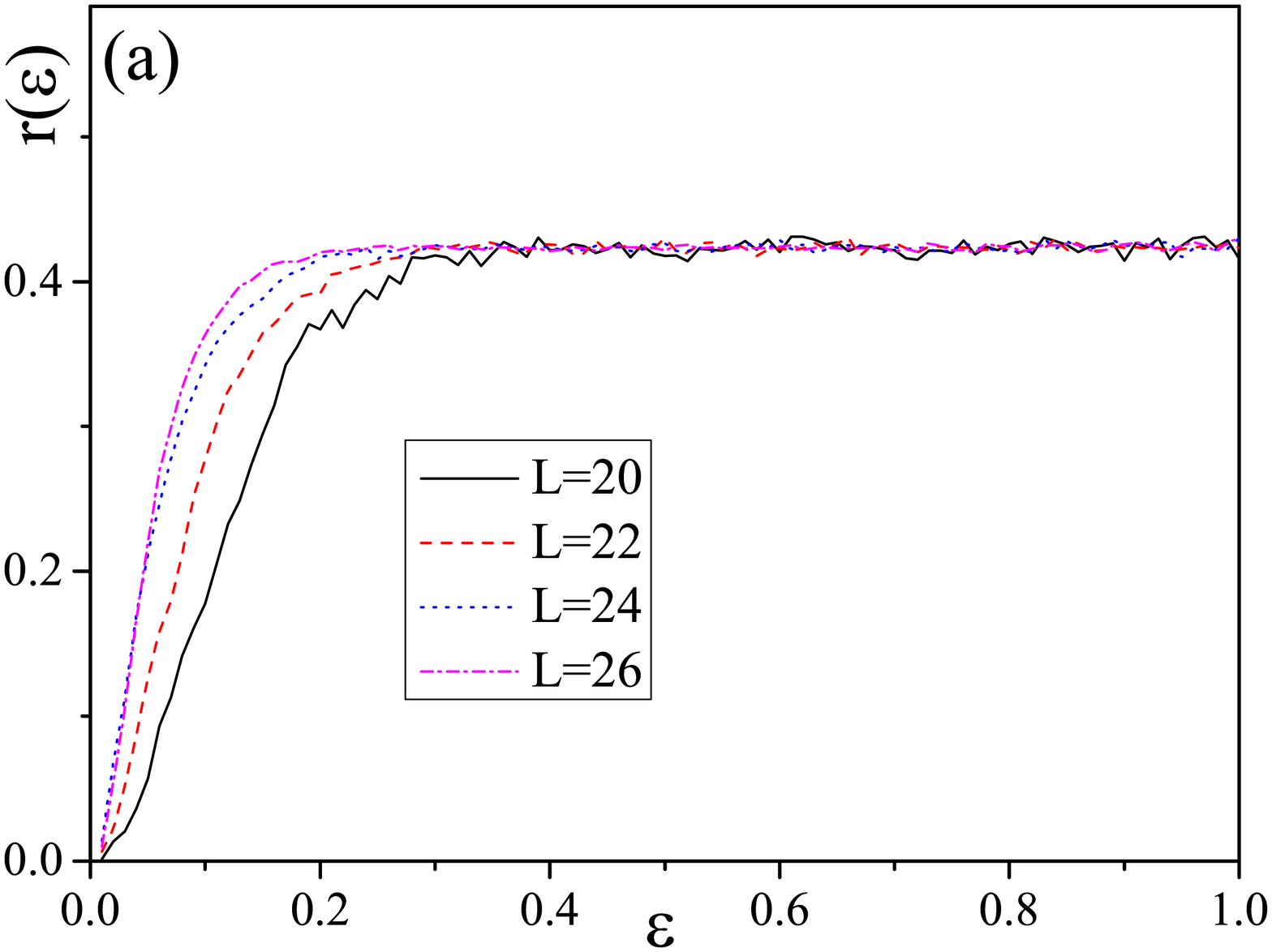}
\includegraphics[width=0.325\linewidth,bb=95 53 760 555]{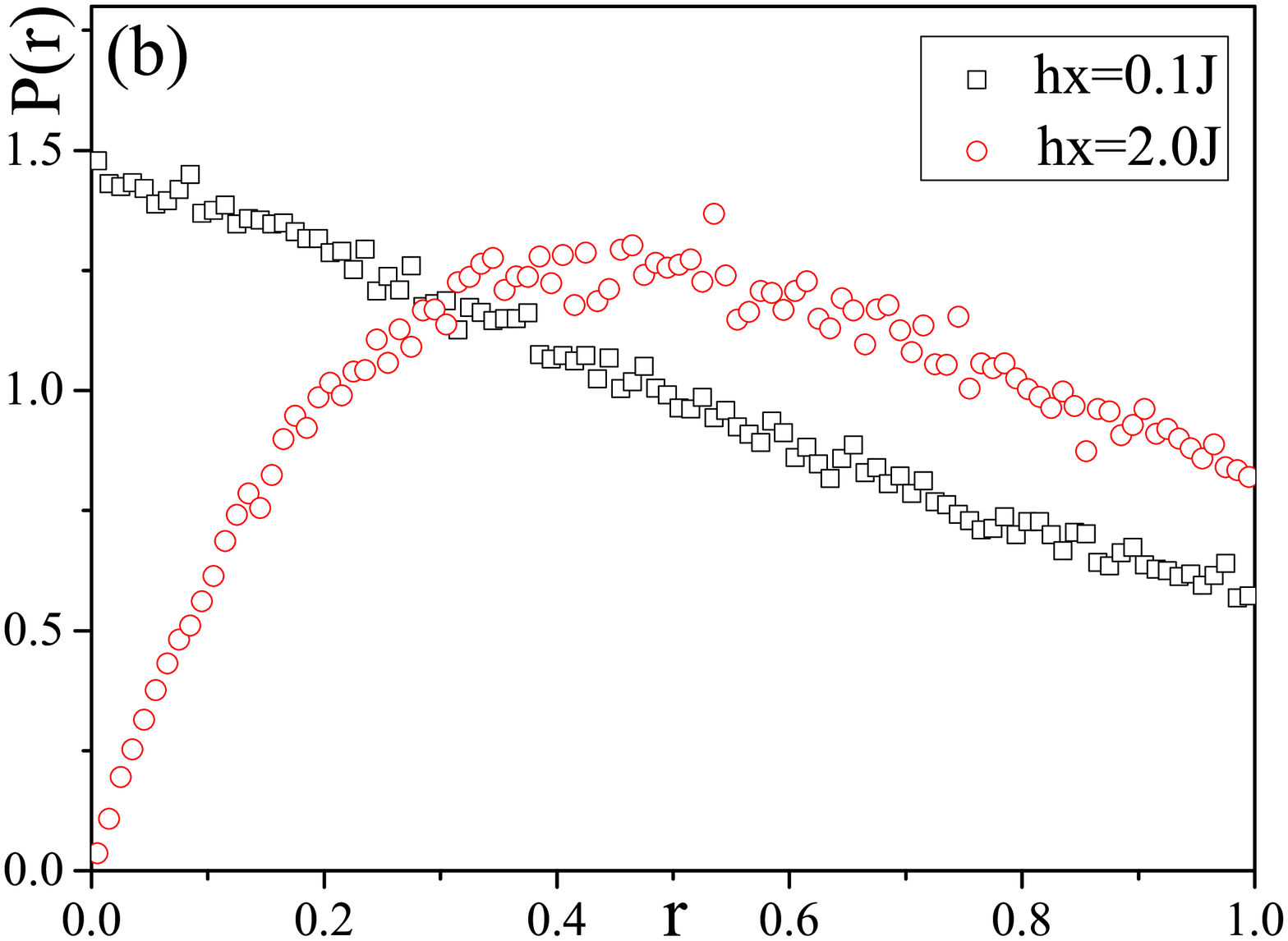}
\includegraphics[width=0.325\linewidth,bb=95 53 760 555]{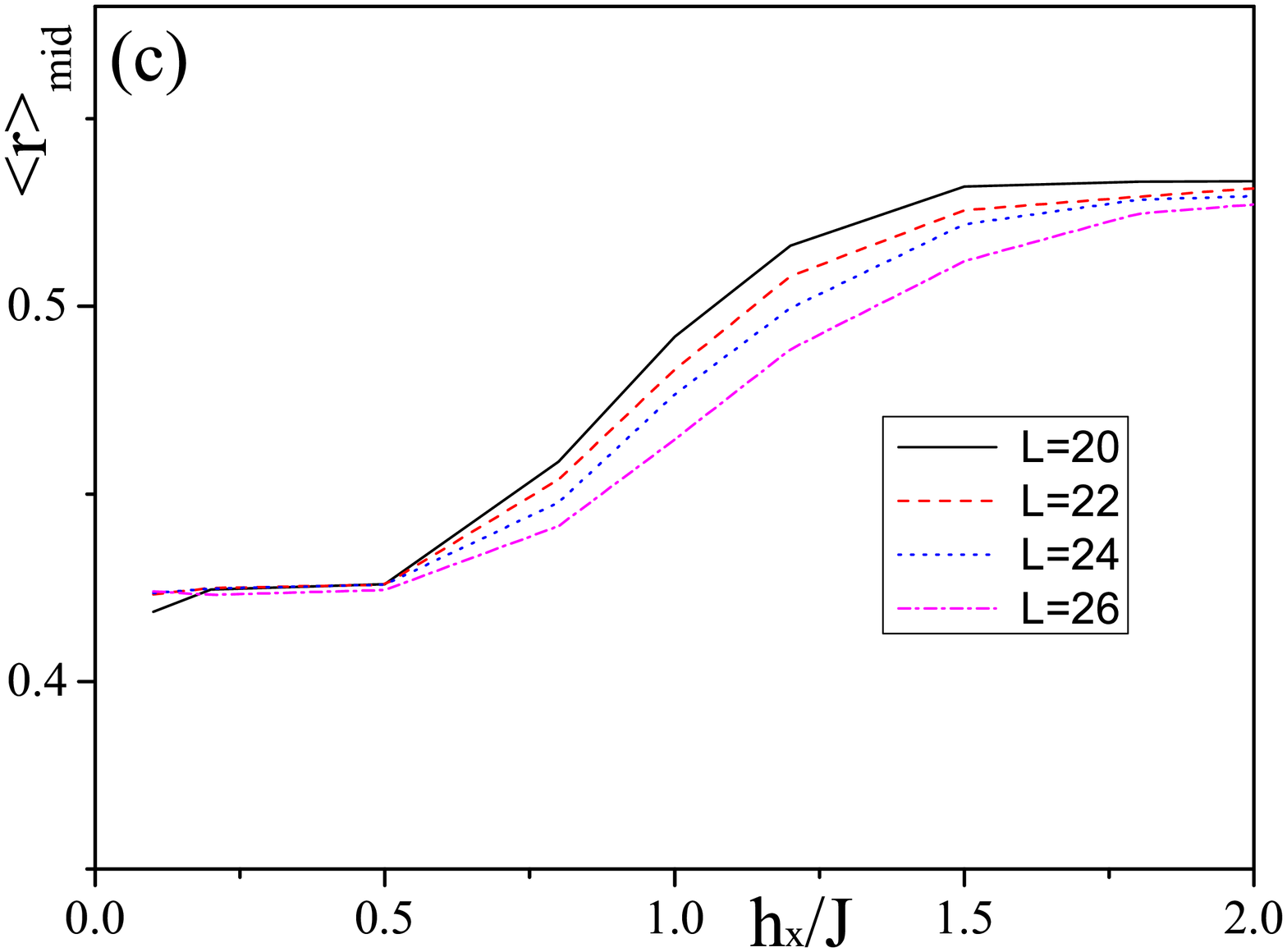}
\includegraphics[width=0.325\linewidth,bb=95 53 760 555]{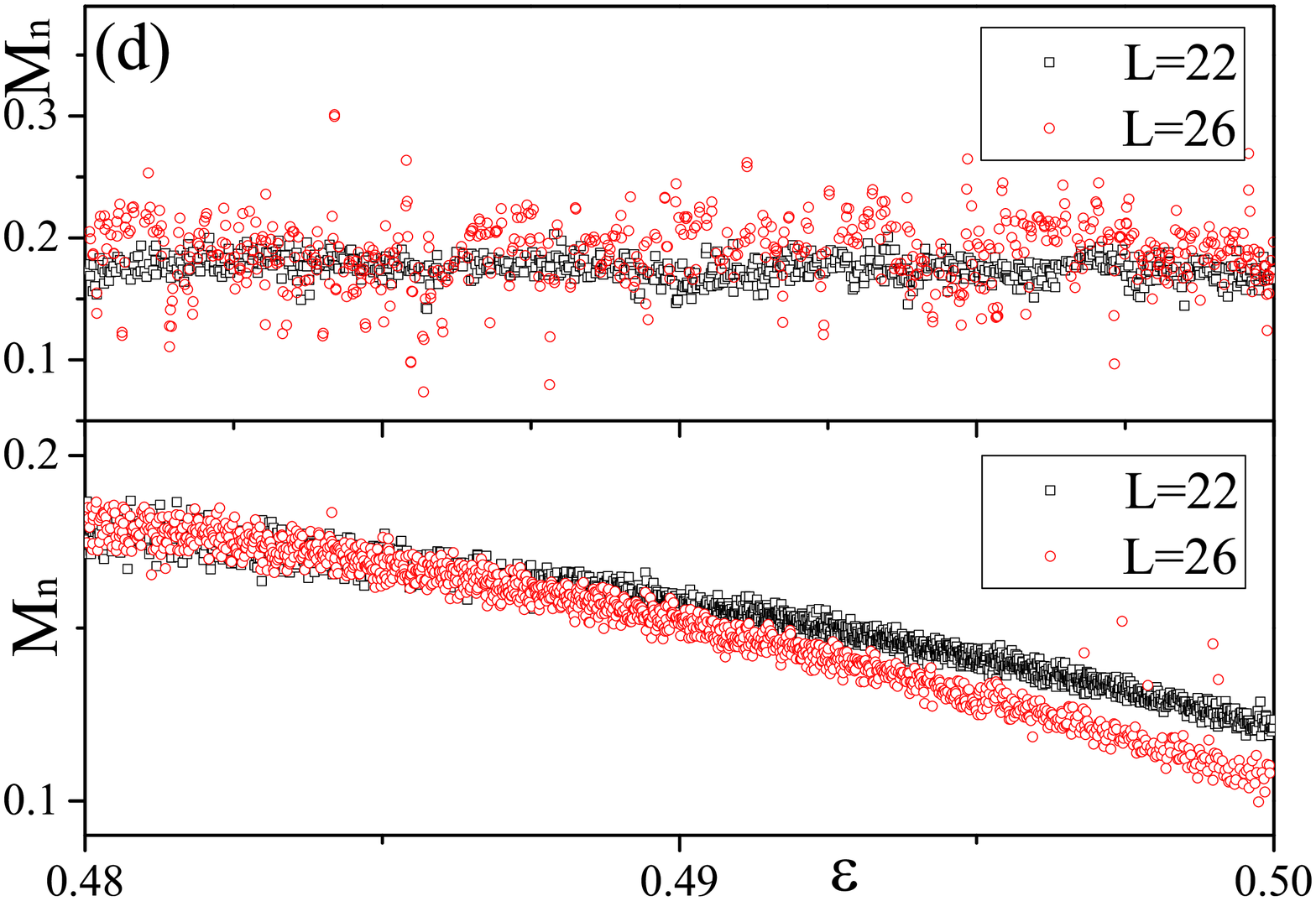}
\includegraphics[width=0.325\linewidth,bb=95 53 760 555]{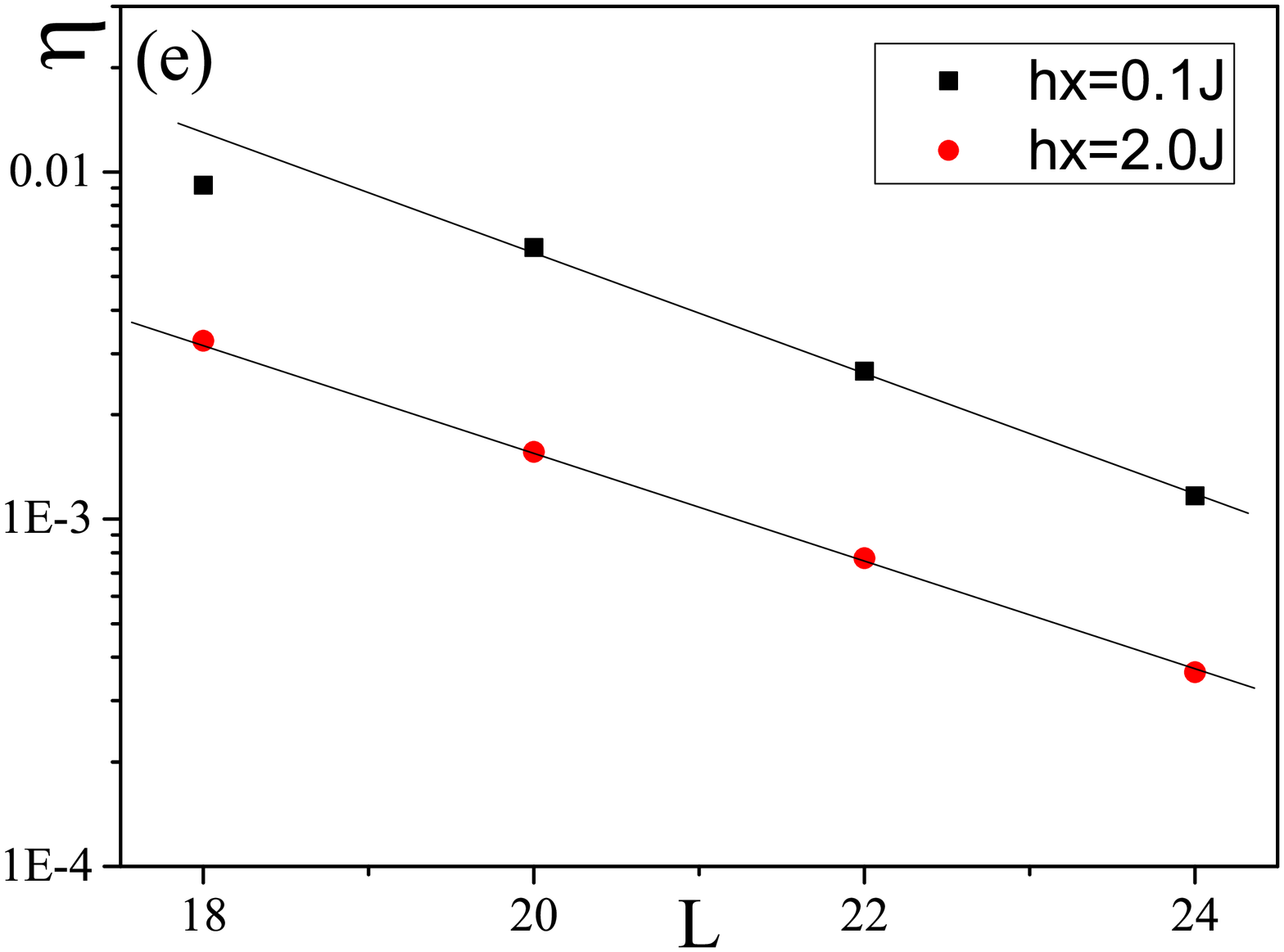}
\includegraphics[width=0.325\linewidth,bb=95 53 760 555]{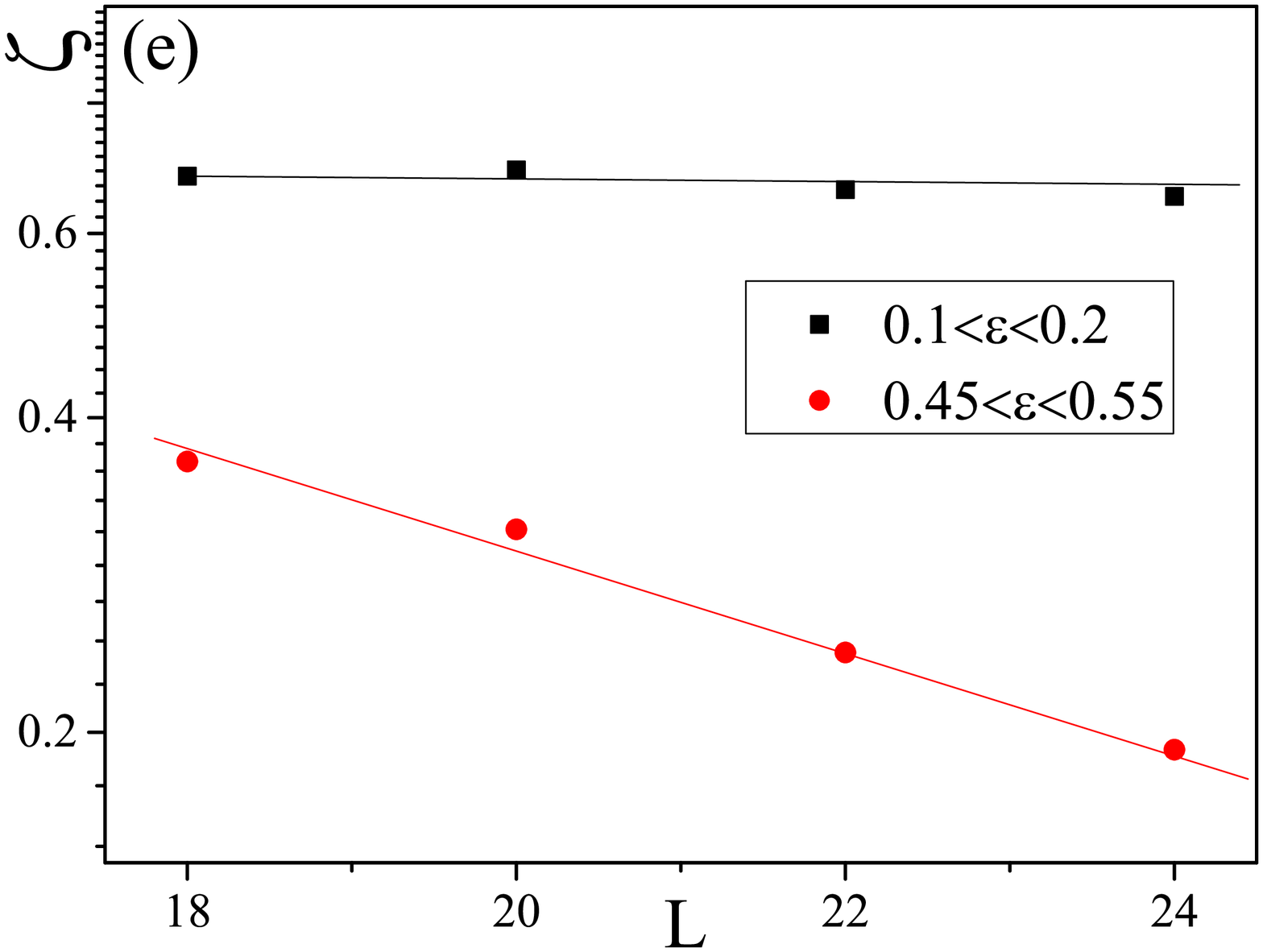}
\caption{(Color online) (a) Average ratio of adjacent level spacing $r(\epsilon)$ as a function of normalized energy $\epsilon$ with different system sizes and fixed $h_x=0.1J$. (b) Distributions of $r_n$ in the ergodic ($h_x=2J$) and nonergodic ($h_x=0.1J$) paramagnetic eigenstates. (c)  Dependence of the average $r_n$ on $h_x$ in the paramagnetic phases. (d) Eigenstate expectation values of the FM order parameters $M_n$ as a function of $\epsilon$ in the nonergodic  ($h_x=0.1J$ upper panel) and ergodic ($h_x=2J$ lower panel) phases. (e) System size dependence of the averaged partition ratio $\eta$ for different $h_x$. (f)System size dependence of the ratio between the average inter- and intracluster spin (glass) correlations in the paramagnetic phase (red circle) and SG phase (black box) with a fixed $h_x=0.1J$. The statistics and average are performed over the eigenstates within the normalized energy window $\epsilon\in [0.45,0.55]$ for (b) (c) and (e). $\mathcal{N}=1000$ for all figures except (d), where no ensemble average is performed ($\mathcal{N}=1$).
} \label{fig:fig3}
\end{figure*}

{\it Spin-glass transition  --} In the classical limit ($h_x=0$), the system experiences a thermal phase transition from a low-temperature SG-like magnetic phase to a high temperature paramagnetic phase. In the presence of a weak transverse field, we conjecture that there may be a similar transition that separates the low- and high-energy eigenstates of the Hamiltonian.(\ref{eq:Ham1}). To verify this point numerically, we define the SG order parameter for the {\it n}-th eigenstate $|n\rangle$\cite{Kjall2014,Mukherjee2018}:
\begin{equation}
\chi_n=\frac 1N \sum_{i,j} \langle n|\hat{s}_i^z\hat{s}_j^z|n\rangle^2
\end{equation}
and calculate its dependence on the normalized energy density $\epsilon=\frac {E_n-E_{min}}{E_{max}-E_{min}}$ where $E_n$ is the eigenenergy of $|n\rangle$, and $E_{min}$ ($E_{max}$) is the minimum (maximum) eigenenergy. We define $\chi(\epsilon)=\langle \chi_n\rangle_\epsilon $, where the average $\langle \rangle_\epsilon$ is performed over all the eigenstates within the energy window $[\epsilon,\epsilon+\Delta]$ with $\Delta=0.01$.  We plot $\chi(\epsilon)$ with a small fixed transverse field ($h_x=0.1J$) for various system sizes in Fig.\ref{fig:fig2}, which shows that at low energy, $\chi(\epsilon)$ linearly diverges with the system size ($\chi(\epsilon)\sim N$): a signature of spin freezing and long-range correlation. In contrast, at high energy,  $\chi(\epsilon)$ approaches a finite value in the thermodynamic limit, indicating a short-range correlation only. The distinct behaviors of $\chi(\epsilon)$ between the low and high energy eigenstates suggest a phase transition between them, which is characterized by the crossing point ($\epsilon_c$) between the $\chi(\epsilon)$ curves with different $L$.

{\it Paramagnetic states with ergodicity breaking: a level statistics diagnosis  --} Ergodicity can be quantified by the presence of energy level repulsion, which in turn can be determined by the ratio of adjacent level spacing in the energy spectrum $r_n=\frac{min(\delta_n,\delta_{n+1})}{max(\delta_n,\delta_{n+1})}$\cite{Oganesyan2007},  and $\delta_n = E_n-E_{n-1}$ is level spacing  between consecutive energy levels in the ordered list of eigenenergies $\{ E_n\}$ of the Hamiltonian. In ergodic phases, the distribution of the level spacing $P(r)$ is expected to follow  Gaussian orthogonal ensemble(GOE)\cite{Wigner1955}, which is characterized by the vanishing of $P(r)$ for $r\rightarrow 0$ (level repulsion) and a mean value $\langle r_n\rangle \simeq 0.53$, while for nonergodic phases, $P(r)$ typically follows Poisson distribution with a mean value $\langle r_n\rangle \simeq 0.39$.

We first focus on the case with a weak transverse field and study the $\epsilon$-dependence of $r(\epsilon)=\langle r_n\rangle_\epsilon$ with a fixed $h_x=0.1J$. As shown in Fig.\ref{fig:fig3} (a), for small $\epsilon$, $r(\epsilon)$ strongly depends on the system size, and its values are significantly smaller than the average values in the GOE or Poisson distributions. This is because at low energy, an SG phase is accompanied by a spontaneous $Z_2$ symmetry breaking. The typical gap between a low-energy eigenstate and its $Z_2$ symmetric counterpart is exponentially small, thus the average value of $r_n$ is lowered\cite{Huse2013}. For large $\epsilon$, $r(\epsilon)$ approaches the value of $0.41$, slightly higher than the mean value of Poisson distributions. To explore the properties of the high-energy eigenstates, we focus on the eigenstates within energy windows around the spectrum center ($\epsilon_n \in [0.45,0.55]$) and calculate their $r_n$ distribution, which resembles the Poisson, but is far from GOE distribution,  as shown in Fig.\ref{fig:fig3} (b).

One may wonder whether the  absence of level repulsion in the case of small $h_x$ is due to some trivial reasons, for instance, a ``hidden'' symmetry other than PS that can be used to further block diagonalize the Hamiltonian and thus give rise to accidental degeneracy between the energy levels in different blocks. To preclude this possibility, we study the case with a large $h_x$, whose Hamiltonian shares the same symmetries with the small $h_x$ ones. Fig.\ref{fig:fig3} (b) shows that the distribution of $r_n$ in the case of $h_x=2$ follows the GOE statistics, indicating that there are two types of PM states: the ergodic eigenstates of the Hamiltonian.(\ref{eq:Ham1}) with large $h_x$ and the nonergodic ones in the small $h_x$ cases. The difference between them can be characterized by $\langle r_n\rangle_{mid}$, where the average $\langle\rangle_{mid}$ is over  eigenstates within energy windows around the spectrum center ($\epsilon_n \in [0.45,0.55]$). The dependence of $\langle r_n\rangle_{mid}$ on $h_x$ for different system sizes is plotted in Fig.\ref{fig:fig3} (c),  which seems to indicate a crossover instead of a phase transition between the two PM phases.

{\it  Eigenstate thermalization hypothesis and its breaking --} Another diagnostic of  ergodic behavior for quantum systems is the ETH, which states that for a sufficiently large generic quantum many-body system, the expectation value of a few-body operator in an eigenstate of the Hamiltonian is a smooth function of its eigenenergy. To examine the ETH, we choose the operator of the FM order parameter: $M_n=\frac 1N\sqrt{\langle n| (\sum_i \hat{s}_i^z)^2|n\rangle}$ and calculate its expectation value in different eigenstates as a function of eigenenergies. As shown in Fig.\ref{fig:fig3} (d), for a fixed $\epsilon$ close to the spectrum center, the distribution of $M_n$ is diverse in the nonergodic PM phase ($h_x=0.1J$) and its variance increases with the system size, indicating that the ETH is broken. In contrast, for $h_x=2.0J$, the situation is qualitatively different. These results agree with those derived from the energy level statistics.

{\it Spontaneous clustering of nonergodic paramagnetic eigenstates --} Next we study the nature of the nonergodic PM eigenstates. First, to distinguish it from the MBL phases, we calculate  the participation ratio of the eigenstates $\eta=\langle \sum_i |\Psi_n(i)|^4\rangle_{mid}$, where $\Psi_n(i)=\langle n|\vec{s}_i\rangle$ is the coefficient of the $n$th eigenstate projected on the $i$th Fock basis $|\vec{s}_i\rangle$, and $\langle\rangle_{mid}$ is defined above.  Fig.\ref{fig:fig3} (e) shows $\eta$ as a function of system size for two different PM phases, both of which exhibit exponential decays,  indicating that there is no localization in the Fock space for both ergodic and nonergodic PM eigenstates, thus neither of them are MBL states.

To explore the structure of the  nonergodic PM eigenstates, we calculate the inter- and intracluster  SG correlations: $C_{i/e}=\frac{1}{N_{i/e}}\sum_{[ij]\in i/e}\langle n|\hat{s}^z_i \hat{s}^z_j |n\rangle^2$, where the summation $\sum_{[ij]\in i/e}$ is over all $N_{i/e}$ bonds connecting two spins within the same/different clusters, and we further perform the average over the eigenstates satisfying $\epsilon_n\in [0.45,0.55]$. We are more interested in the ratio between the  inter- and intra-cluster  spin correlations: $\zeta=C_e/C_i$. As shown in Fig.\ref{fig:fig3} (f), in the nonergodic PM, $\zeta$ exhibits an exponential decay with  system size, which indicates that in the thermodynamic limit, the high-energy eigenstates are organized into clusters, each of which can be considered as a large quantum spin and has no correlation with others. A single quantum spin (LMG model)  cannot be considered as a generic quantum many-body system, and its energy level space statistics  obey  neither a Poisson nor a GOE distribution (see the SM\cite{Supplementary}), whereas a collection of the energy levels in these decoupled clusters results in the absence of  energy level repulsion. However, for the SG phase, $\zeta$ barely decays with system size, which indicates the different clusters are strongly correlated.

{\it Discussion --} Clustering phenomena have been proposed in the intermediate phase between the SG and PM  phases in the quantum p-spin model\cite{Baldwin2017,Winer2022}, and a delocalized yet nonergodic phase was also observed as an intermediate state in the single-particle disordered system on the Bethe lattice\cite{Luca2014,Burin2017}. Here, the nonergodic phase governs the whole PM  instead of an intermediate regime. Furthermore, unlike the quantum p-spin model where the subsystem is ergodic within each cluster and the total system is partially chaotic, in the quantum HNN model, the PS in each cluster make it become a single quantum object, thus ergodicity is broken even within the clusters. It is an interesting question whether a PS breaking perturbation can immediately destroy the nonergodic phase.

{\it Conclusion and outlook --} In summary, we have studied the high-energy eigenstates of the quantum HNN Hamiltonian, and found a nonergodic yet delocalized paramagnetic states as a consequence of a combination of a spontaneous clustering and the PS in each cluster.  Future developments will include studies of the real-time evolution of this model, including the quantum quench and periodically driven dynamics.  In general, ergodicity breaking indicates that the system will not equilibrate to a thermal state, but whether it will approach a nonthermal steady state or exhibit persistent oscillations like the quantum scar \cite{Turner2018} or the infinite-range interacting systems \cite{Yuzbashyan2006,Barankov2006,Chen2020} is an interesting question worthy of further study. Furthermore, imposing a periodic drive on such a nonergodic model does not necessarily drive the system into an infinite temperature state, and thus may open  new possibilities to explore nontrivial dynamics such as discrete time crystals\cite{Sacha2015,Else2016,Khemani2016,Yao2017}. Finally, since the HNN model is proposed to mimic associative memory, a fundamental question is the relationship between the associative memory and the ergodicity breaking in this quantum model. However, a mimic of the associative memory calls for dissipation which has not been considered here. Incorporating dissipation further complicates the system, but might give rise to intriguing phenomena due to the interplay between the quantum fluctuation and frustration\cite{Rotondo2018,Fiorelli2020}.

{\it Acknowledgments}.---This work is supported by the National Key Research and Development Program of China (Grant No.2020YFA0309000), Natural Science Foundation of  China (Grant No.12174251),  Natural Science Foundation of Shanghai (Grant No.22ZR142830),   Shanghai Municipal Science and Technology Major Project (Grant No.2019SHZDZX01).


\begin{thebibliography}{40}
\expandafter\ifx\csname natexlab\endcsname\relax\def\natexlab#1{#1}\fi
\expandafter\ifx\csname bibnamefont\endcsname\relax
  \def\bibnamefont#1{#1}\fi
\expandafter\ifx\csname bibfnamefont\endcsname\relax
  \def\bibfnamefont#1{#1}\fi
\expandafter\ifx\csname citenamefont\endcsname\relax
  \def\citenamefont#1{#1}\fi
\expandafter\ifx\csname url\endcsname\relax
  \def\url#1{\texttt{#1}}\fi
\expandafter\ifx\csname urlprefix\endcsname\relax\def\urlprefix{URL }\fi
\providecommand{\bibinfo}[2]{#2}
\providecommand{\eprint}[2][]{\url{#2}}

\bibitem[{\citenamefont{Rigol et~al.}(2008)\citenamefont{Rigol, Dunjko, and
  Olshanii}}]{Rigol2008}
\bibinfo{author}{\bibfnamefont{M.}~\bibnamefont{Rigol}},
  \bibinfo{author}{\bibfnamefont{V.}~\bibnamefont{Dunjko}}, \bibnamefont{and}
  \bibinfo{author}{\bibfnamefont{M.}~\bibnamefont{Olshanii}},
  \bibinfo{journal}{Nature} \textbf{\bibinfo{volume}{452}},
  \bibinfo{pages}{854} (\bibinfo{year}{2008}).

\bibitem[{\citenamefont{D.M.Basko et~al.}(2006)\citenamefont{D.M.Basko,
  I.L.Aleiner, and B.L.Altshuler}}]{Basko2006}
\bibinfo{author}{\bibnamefont{D.M.Basko}},
  \bibinfo{author}{\bibnamefont{I.L.Aleiner}}, \bibnamefont{and}
  \bibinfo{author}{\bibnamefont{B.L.Altshuler}}, \bibinfo{journal}{Annals of
  Physics} \textbf{\bibinfo{volume}{321}}, \bibinfo{pages}{1126}
  (\bibinfo{year}{2006}).

\bibitem[{\citenamefont{Oganesyan and Huse}(2007)}]{Oganesyan2007}
\bibinfo{author}{\bibfnamefont{V.}~\bibnamefont{Oganesyan}} \bibnamefont{and}
  \bibinfo{author}{\bibfnamefont{D.~A.} \bibnamefont{Huse}},
  \bibinfo{journal}{Phys. Rev. B} \textbf{\bibinfo{volume}{75}},
  \bibinfo{pages}{155111} (\bibinfo{year}{2007}).

\bibitem[{\citenamefont{\ifmmode \check{Z}\else
  \v{Z}\fi{}nidari\ifmmode~\check{c}\else \v{c}\fi{}
  et~al.}(2008)\citenamefont{\ifmmode \check{Z}\else
  \v{Z}\fi{}nidari\ifmmode~\check{c}\else \v{c}\fi{}, Prosen, and
  Prelov\ifmmode~\check{s}\else \v{s}\fi{}ek}}]{Znidaric2008}
\bibinfo{author}{\bibfnamefont{M.}~\bibnamefont{\ifmmode \check{Z}\else
  \v{Z}\fi{}nidari\ifmmode~\check{c}\else \v{c}\fi{}}},
  \bibinfo{author}{\bibfnamefont{T.}~\bibnamefont{Prosen}}, \bibnamefont{and}
  \bibinfo{author}{\bibfnamefont{P.}~\bibnamefont{Prelov\ifmmode~\check{s}\else
  \v{s}\fi{}ek}}, \bibinfo{journal}{Phys. Rev. B}
  \textbf{\bibinfo{volume}{77}}, \bibinfo{pages}{064426}
  (\bibinfo{year}{2008}).

\bibitem[{\citenamefont{Pal and Huse}(2010)}]{Pal2010}
\bibinfo{author}{\bibfnamefont{A.}~\bibnamefont{Pal}} \bibnamefont{and}
  \bibinfo{author}{\bibfnamefont{D.~A.} \bibnamefont{Huse}},
  \bibinfo{journal}{Phys. Rev. B} \textbf{\bibinfo{volume}{82}},
  \bibinfo{pages}{174411} (\bibinfo{year}{2010}).

\bibitem[{\citenamefont{Mezard et~al.}(1986)\citenamefont{Mezard, Parisi, and
  Virasoro}}]{Mezard1986}
\bibinfo{author}{\bibfnamefont{M.}~\bibnamefont{Mezard}},
  \bibinfo{author}{\bibfnamefont{G.}~\bibnamefont{Parisi}}, \bibnamefont{and}
  \bibinfo{author}{\bibfnamefont{M.}~\bibnamefont{Virasoro}},
  \emph{\bibinfo{title}{Spin Glass Theory and Beyond: An Introduction to the
  Replica Method and Its Applications}} (\bibinfo{publisher}{~World
  Scientific}, \bibinfo{year}{1986}).

\bibitem[{\citenamefont{Rigol et~al.}(2007)\citenamefont{Rigol, Dunjko,
  Yurovsky, and Olshanii}}]{Rigol2007}
\bibinfo{author}{\bibfnamefont{M.}~\bibnamefont{Rigol}},
  \bibinfo{author}{\bibfnamefont{V.}~\bibnamefont{Dunjko}},
  \bibinfo{author}{\bibfnamefont{V.}~\bibnamefont{Yurovsky}}, \bibnamefont{and}
  \bibinfo{author}{\bibfnamefont{M.}~\bibnamefont{Olshanii}},
  \bibinfo{journal}{Phys. Rev. Lett.} \textbf{\bibinfo{volume}{98}},
  \bibinfo{pages}{050405} (\bibinfo{year}{2007}).

\bibitem[{\citenamefont{Turner et~al.}(2018)\citenamefont{Turner, Michailidis,
  Abanin, Serbyn, and Papic}}]{Turner2018}
\bibinfo{author}{\bibfnamefont{C.~J.} \bibnamefont{Turner}},
  \bibinfo{author}{\bibfnamefont{A.~A.} \bibnamefont{Michailidis}},
  \bibinfo{author}{\bibfnamefont{D.~A.} \bibnamefont{Abanin}},
  \bibinfo{author}{\bibfnamefont{M.}~\bibnamefont{Serbyn}}, \bibnamefont{and}
  \bibinfo{author}{\bibfnamefont{Z.}~\bibnamefont{Papic}},
  \bibinfo{journal}{Nat Phys} \textbf{\bibinfo{volume}{14}},
  \bibinfo{pages}{745} (\bibinfo{year}{2018}).

\bibitem[{\citenamefont{Edwards and Anderson}(1975)}]{Edwards1975}
\bibinfo{author}{\bibfnamefont{S.~F.} \bibnamefont{Edwards}} \bibnamefont{and}
  \bibinfo{author}{\bibfnamefont{P.~W.} \bibnamefont{Anderson}},
  \bibinfo{journal}{J. Phys. F} \textbf{\bibinfo{volume}{5}},
  \bibinfo{pages}{965} (\bibinfo{year}{1975}).

\bibitem[{\citenamefont{Baldwin et~al.}(2017)\citenamefont{Baldwin, Laumann,
  Pal, and Scardicchio}}]{Baldwin2017}
\bibinfo{author}{\bibfnamefont{C.~L.} \bibnamefont{Baldwin}},
  \bibinfo{author}{\bibfnamefont{C.~R.} \bibnamefont{Laumann}},
  \bibinfo{author}{\bibfnamefont{A.}~\bibnamefont{Pal}}, \bibnamefont{and}
  \bibinfo{author}{\bibfnamefont{A.}~\bibnamefont{Scardicchio}},
  \bibinfo{journal}{Phys. Rev. Lett.} \textbf{\bibinfo{volume}{118}},
  \bibinfo{pages}{127201} (\bibinfo{year}{2017}).

\bibitem[{\citenamefont{Mukherjee et~al.}(2018)\citenamefont{Mukherjee, Nag,
  and Garg}}]{Mukherjee2018}
\bibinfo{author}{\bibfnamefont{S.}~\bibnamefont{Mukherjee}},
  \bibinfo{author}{\bibfnamefont{S.}~\bibnamefont{Nag}}, \bibnamefont{and}
  \bibinfo{author}{\bibfnamefont{A.}~\bibnamefont{Garg}},
  \bibinfo{journal}{Phys. Rev. B} \textbf{\bibinfo{volume}{97}},
  \bibinfo{pages}{144202} (\bibinfo{year}{2018}).

\bibitem[{\citenamefont{Rademaker and Abanin}(2020)}]{Rademaker2020}
\bibinfo{author}{\bibfnamefont{L.}~\bibnamefont{Rademaker}} \bibnamefont{and}
  \bibinfo{author}{\bibfnamefont{D.~A.} \bibnamefont{Abanin}},
  \bibinfo{journal}{Phys. Rev. Lett.} \textbf{\bibinfo{volume}{125}},
  \bibinfo{pages}{260405} (\bibinfo{year}{2020}).

\bibitem[{\citenamefont{Thomson et~al.}(2020)\citenamefont{Thomson, Urbani, and
  Schir\'o}}]{Thomson2020}
\bibinfo{author}{\bibfnamefont{S.~J.} \bibnamefont{Thomson}},
  \bibinfo{author}{\bibfnamefont{P.}~\bibnamefont{Urbani}}, \bibnamefont{and}
  \bibinfo{author}{\bibfnamefont{M.}~\bibnamefont{Schir\'o}},
  \bibinfo{journal}{Phys. Rev. Lett.} \textbf{\bibinfo{volume}{125}},
  \bibinfo{pages}{120602} (\bibinfo{year}{2020}).

\bibitem[{\citenamefont{{Winer} et~al.}(2022)\citenamefont{{Winer}, {Barney},
  {Baldwin}, {Galitski}, and {Swingle}}}]{Winer2022}
\bibinfo{author}{\bibfnamefont{M.}~\bibnamefont{{Winer}}},
  \bibinfo{author}{\bibfnamefont{R.}~\bibnamefont{{Barney}}},
  \bibinfo{author}{\bibfnamefont{C.~L.} \bibnamefont{{Baldwin}}},
  \bibinfo{author}{\bibfnamefont{V.}~\bibnamefont{{Galitski}}},
  \bibnamefont{and}
  \bibinfo{author}{\bibfnamefont{B.}~\bibnamefont{{Swingle}}},
  \bibinfo{journal}{arXiv e-prints} \bibinfo{eid}{arXiv:2203.12753}
  (\bibinfo{year}{2022}), \eprint{2203.12753}.

\bibitem[{\citenamefont{Hopfield}(1982)}]{Hopfield1982}
\bibinfo{author}{\bibfnamefont{J.~J.} \bibnamefont{Hopfield}},
  \bibinfo{journal}{PNAS} \textbf{\bibinfo{volume}{79}}, \bibinfo{pages}{2554}
  (\bibinfo{year}{1982}).

\bibitem[{\citenamefont{Amit et~al.}(1985{\natexlab{a}})\citenamefont{Amit,
  Gutfreund, and Sompolinsky}}]{Amit1985a}
\bibinfo{author}{\bibfnamefont{D.~J.} \bibnamefont{Amit}},
  \bibinfo{author}{\bibfnamefont{H.}~\bibnamefont{Gutfreund}},
  \bibnamefont{and}
  \bibinfo{author}{\bibfnamefont{H.}~\bibnamefont{Sompolinsky}},
  \bibinfo{journal}{Phys. Rev. Lett.} \textbf{\bibinfo{volume}{55}},
  \bibinfo{pages}{1530} (\bibinfo{year}{1985}{\natexlab{a}}).

\bibitem[{\citenamefont{Amit et~al.}(1985{\natexlab{b}})\citenamefont{Amit,
  Gutfreund, and Sompolinsky}}]{Amit1985b}
\bibinfo{author}{\bibfnamefont{D.~J.} \bibnamefont{Amit}},
  \bibinfo{author}{\bibfnamefont{H.}~\bibnamefont{Gutfreund}},
  \bibnamefont{and}
  \bibinfo{author}{\bibfnamefont{H.}~\bibnamefont{Sompolinsky}},
  \bibinfo{journal}{Phys. Rev. A} \textbf{\bibinfo{volume}{32}},
  \bibinfo{pages}{1007} (\bibinfo{year}{1985}{\natexlab{b}}).

\bibitem[{\citenamefont{van Hemmen}(1982)}]{Hemmen1982}
\bibinfo{author}{\bibfnamefont{J.~L.} \bibnamefont{van Hemmen}},
  \bibinfo{journal}{Phys. Rev. Lett.} \textbf{\bibinfo{volume}{49}},
  \bibinfo{pages}{409} (\bibinfo{year}{1982}).

\bibitem[{\citenamefont{Rotondo et~al.}(2018)\citenamefont{Rotondo, Marcuzzi,
  Garrahan, I.Lesanovsky, and Muller}}]{Rotondo2018}
\bibinfo{author}{\bibfnamefont{P.}~\bibnamefont{Rotondo}},
  \bibinfo{author}{\bibfnamefont{M.}~\bibnamefont{Marcuzzi}},
  \bibinfo{author}{\bibfnamefont{J.}~\bibnamefont{Garrahan}},
  \bibinfo{author}{\bibnamefont{I.Lesanovsky}}, \bibnamefont{and}
  \bibinfo{author}{\bibfnamefont{M.}~\bibnamefont{Muller}},
  \bibinfo{journal}{J. Phys. A} \textbf{\bibinfo{volume}{51}},
  \bibinfo{pages}{115301} (\bibinfo{year}{2018}).

\bibitem[{\citenamefont{Fiorelli et~al.}(2020)\citenamefont{Fiorelli, Marcuzzi,
  Rotondo, Carollo, and Lesanovsky}}]{Fiorelli2020}
\bibinfo{author}{\bibfnamefont{E.}~\bibnamefont{Fiorelli}},
  \bibinfo{author}{\bibfnamefont{M.}~\bibnamefont{Marcuzzi}},
  \bibinfo{author}{\bibfnamefont{P.}~\bibnamefont{Rotondo}},
  \bibinfo{author}{\bibfnamefont{F.}~\bibnamefont{Carollo}}, \bibnamefont{and}
  \bibinfo{author}{\bibfnamefont{I.}~\bibnamefont{Lesanovsky}},
  \bibinfo{journal}{Phys. Rev. Lett.} \textbf{\bibinfo{volume}{125}},
  \bibinfo{pages}{070604} (\bibinfo{year}{2020}).

\bibitem[{\citenamefont{Deutsch}(1991)}]{Deutsch1991}
\bibinfo{author}{\bibfnamefont{J.~M.} \bibnamefont{Deutsch}},
  \bibinfo{journal}{Phys. Rev. A} \textbf{\bibinfo{volume}{43}},
  \bibinfo{pages}{2046} (\bibinfo{year}{1991}).

\bibitem[{\citenamefont{Srednicki}(1994)}]{Srednicki1994}
\bibinfo{author}{\bibfnamefont{M.}~\bibnamefont{Srednicki}},
  \bibinfo{journal}{Phys. Rev. E} \textbf{\bibinfo{volume}{50}},
  \bibinfo{pages}{888} (\bibinfo{year}{1994}).

\bibitem[{\citenamefont{Fratus and Srednicki}(2015)}]{Fratus2015}
\bibinfo{author}{\bibfnamefont{K.~R.} \bibnamefont{Fratus}} \bibnamefont{and}
  \bibinfo{author}{\bibfnamefont{M.}~\bibnamefont{Srednicki}},
  \bibinfo{journal}{Phys. Rev. E} \textbf{\bibinfo{volume}{92}},
  \bibinfo{pages}{040103} (\bibinfo{year}{2015}).

\bibitem[{\citenamefont{Laumann et~al.}(2014)\citenamefont{Laumann, Pal, and
  Scardicchio}}]{Laumann2014}
\bibinfo{author}{\bibfnamefont{C.~R.} \bibnamefont{Laumann}},
  \bibinfo{author}{\bibfnamefont{A.}~\bibnamefont{Pal}}, \bibnamefont{and}
  \bibinfo{author}{\bibfnamefont{A.}~\bibnamefont{Scardicchio}},
  \bibinfo{journal}{Phys. Rev. Lett.} \textbf{\bibinfo{volume}{113}},
  \bibinfo{pages}{200405} (\bibinfo{year}{2014}).

\bibitem[{\citenamefont{Mattis}(1976)}]{Mattis1976}
\bibinfo{author}{\bibfnamefont{D.~C.} \bibnamefont{Mattis}},
  \bibinfo{journal}{Phys. Lett. A} \textbf{\bibinfo{volume}{56}},
  \bibinfo{pages}{421} (\bibinfo{year}{1976}).

\bibitem[{\citenamefont{Carollo and Lesanovsky}(2021)}]{Carollo2021}
\bibinfo{author}{\bibfnamefont{F.}~\bibnamefont{Carollo}} \bibnamefont{and}
  \bibinfo{author}{\bibfnamefont{I.}~\bibnamefont{Lesanovsky}},
  \bibinfo{journal}{Phys. Rev. Lett.} \textbf{\bibinfo{volume}{126}},
  \bibinfo{pages}{230601} (\bibinfo{year}{2021}).

\bibitem[{\citenamefont{Lipkin et~al.}(1965)\citenamefont{Lipkin, Meshkov, and
  Glick}}]{Lipkin1965}
\bibinfo{author}{\bibfnamefont{H.~J.} \bibnamefont{Lipkin}},
  \bibinfo{author}{\bibfnamefont{N.}~\bibnamefont{Meshkov}}, \bibnamefont{and}
  \bibinfo{author}{\bibfnamefont{A.~J.} \bibnamefont{Glick}},
  \bibinfo{journal}{Nucl. Phys.} \textbf{\bibinfo{volume}{62}},
  \bibinfo{pages}{188} (\bibinfo{year}{1965}).

\bibitem[{Sup()}]{Supplementary}
\bibinfo{howpublished}{See the supplementary material for the classification
  scheme and the effective Hamiltonian of the $p=4$ case, a discussion of the
  possibility of special memory patterns with additional symmetries, and an
  analysis of the energy level statistics of the Lipkin-Meshkov-Glick model.}

\bibitem[{\citenamefont{Kj\"all et~al.}(2014)\citenamefont{Kj\"all, Bardarson,
  and Pollmann}}]{Kjall2014}
\bibinfo{author}{\bibfnamefont{J.~A.} \bibnamefont{Kj\"all}},
  \bibinfo{author}{\bibfnamefont{J.~H.} \bibnamefont{Bardarson}},
  \bibnamefont{and} \bibinfo{author}{\bibfnamefont{F.}~\bibnamefont{Pollmann}},
  \bibinfo{journal}{Phys. Rev. Lett.} \textbf{\bibinfo{volume}{113}},
  \bibinfo{pages}{107204} (\bibinfo{year}{2014}).

\bibitem[{\citenamefont{Wigner}(1955)}]{Wigner1955}
\bibinfo{author}{\bibfnamefont{E.~P.} \bibnamefont{Wigner}},
  \bibinfo{journal}{Annals of Mathematics} \textbf{\bibinfo{volume}{61}},
  \bibinfo{pages}{548} (\bibinfo{year}{1955}).

\bibitem[{\citenamefont{Huse et~al.}(2013)\citenamefont{Huse, Nandkishore,
  Oganesyan, Pal, and Sondhi}}]{Huse2013}
\bibinfo{author}{\bibfnamefont{D.~A.} \bibnamefont{Huse}},
  \bibinfo{author}{\bibfnamefont{R.}~\bibnamefont{Nandkishore}},
  \bibinfo{author}{\bibfnamefont{V.}~\bibnamefont{Oganesyan}},
  \bibinfo{author}{\bibfnamefont{A.}~\bibnamefont{Pal}}, \bibnamefont{and}
  \bibinfo{author}{\bibfnamefont{S.~L.} \bibnamefont{Sondhi}},
  \bibinfo{journal}{Phys. Rev. B} \textbf{\bibinfo{volume}{88}},
  \bibinfo{pages}{014206} (\bibinfo{year}{2013}).

\bibitem[{\citenamefont{De~Luca et~al.}(2014)\citenamefont{De~Luca, Altshuler,
  Kravtsov, and Scardicchio}}]{Luca2014}
\bibinfo{author}{\bibfnamefont{A.}~\bibnamefont{De~Luca}},
  \bibinfo{author}{\bibfnamefont{B.~L.} \bibnamefont{Altshuler}},
  \bibinfo{author}{\bibfnamefont{V.~E.} \bibnamefont{Kravtsov}},
  \bibnamefont{and}
  \bibinfo{author}{\bibfnamefont{A.}~\bibnamefont{Scardicchio}},
  \bibinfo{journal}{Phys. Rev. Lett.} \textbf{\bibinfo{volume}{113}},
  \bibinfo{pages}{046806} (\bibinfo{year}{2014}).

\bibitem[{\citenamefont{Burin}(2017)}]{Burin2017}
\bibinfo{author}{\bibfnamefont{A.}~\bibnamefont{Burin}}, \bibinfo{journal}{Ann.
  Phys.(Amsterdam)} \textbf{\bibinfo{volume}{529}}, \bibinfo{pages}{1600292}
  (\bibinfo{year}{2017}).

\bibitem[{\citenamefont{Yuzbashyan et~al.}(2006)\citenamefont{Yuzbashyan,
  Tsyplyatyev, and Altshuler}}]{Yuzbashyan2006}
\bibinfo{author}{\bibfnamefont{E.~A.} \bibnamefont{Yuzbashyan}},
  \bibinfo{author}{\bibfnamefont{O.}~\bibnamefont{Tsyplyatyev}},
  \bibnamefont{and} \bibinfo{author}{\bibfnamefont{B.~L.}
  \bibnamefont{Altshuler}}, \bibinfo{journal}{Phys. Rev. Lett.}
  \textbf{\bibinfo{volume}{96}}, \bibinfo{pages}{097005}
  (\bibinfo{year}{2006}).

\bibitem[{\citenamefont{Barankov and Levitov}(2006)}]{Barankov2006}
\bibinfo{author}{\bibfnamefont{R.~A.} \bibnamefont{Barankov}} \bibnamefont{and}
  \bibinfo{author}{\bibfnamefont{L.~S.} \bibnamefont{Levitov}},
  \bibinfo{journal}{Phys. Rev. Lett.} \textbf{\bibinfo{volume}{96}},
  \bibinfo{pages}{230403} (\bibinfo{year}{2006}).

\bibitem[{\citenamefont{Chen and Cai}(2020)}]{Chen2020}
\bibinfo{author}{\bibfnamefont{Y.}~\bibnamefont{Chen}} \bibnamefont{and}
  \bibinfo{author}{\bibfnamefont{Z.}~\bibnamefont{Cai}},
  \bibinfo{journal}{Phys. Rev. A} \textbf{\bibinfo{volume}{101}},
  \bibinfo{pages}{023611} (\bibinfo{year}{2020}).

\bibitem[{\citenamefont{Sacha}(2015)}]{Sacha2015}
\bibinfo{author}{\bibfnamefont{K.}~\bibnamefont{Sacha}},
  \bibinfo{journal}{Phys. Rev. A} \textbf{\bibinfo{volume}{91}},
  \bibinfo{pages}{033617} (\bibinfo{year}{2015}).

\bibitem[{\citenamefont{Else et~al.}(2016)\citenamefont{Else, Bauer, and
  Nayak}}]{Else2016}
\bibinfo{author}{\bibfnamefont{D.~V.} \bibnamefont{Else}},
  \bibinfo{author}{\bibfnamefont{B.}~\bibnamefont{Bauer}}, \bibnamefont{and}
  \bibinfo{author}{\bibfnamefont{C.}~\bibnamefont{Nayak}},
  \bibinfo{journal}{Phys. Rev. Lett.} \textbf{\bibinfo{volume}{117}},
  \bibinfo{pages}{090402} (\bibinfo{year}{2016}).

\bibitem[{\citenamefont{Khemani et~al.}(2016)\citenamefont{Khemani, Lazarides,
  Moessner, and Sondhi}}]{Khemani2016}
\bibinfo{author}{\bibfnamefont{V.}~\bibnamefont{Khemani}},
  \bibinfo{author}{\bibfnamefont{A.}~\bibnamefont{Lazarides}},
  \bibinfo{author}{\bibfnamefont{R.}~\bibnamefont{Moessner}}, \bibnamefont{and}
  \bibinfo{author}{\bibfnamefont{S.~L.} \bibnamefont{Sondhi}},
  \bibinfo{journal}{Phys. Rev. Lett.} \textbf{\bibinfo{volume}{116}},
  \bibinfo{pages}{250401} (\bibinfo{year}{2016}).

\bibitem[{\citenamefont{Yao et~al.}(2017)\citenamefont{Yao, Potter, Potirniche,
  and Vishwanath}}]{Yao2017}
\bibinfo{author}{\bibfnamefont{N.~Y.} \bibnamefont{Yao}},
  \bibinfo{author}{\bibfnamefont{A.~C.} \bibnamefont{Potter}},
  \bibinfo{author}{\bibfnamefont{I.-D.} \bibnamefont{Potirniche}},
  \bibnamefont{and}
  \bibinfo{author}{\bibfnamefont{A.}~\bibnamefont{Vishwanath}},
  \bibinfo{journal}{Phys. Rev. Lett.} \textbf{\bibinfo{volume}{118}},
  \bibinfo{pages}{030401} (\bibinfo{year}{2017}).

\end{thebibliography}

\end{document}